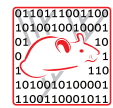

RESEARCH  Open Access

# PAV ontology: provenance, authoring and versioning

Paolo Ciccarese[1,2*†], Stian Soiland-Reyes[3†], Khalid Belhajjame[3], Alasdair JG Gray[3], Carole Goble[3] and Tim Clark[1,2,3]

**Abstract**

**Background:** Provenance is a critical ingredient for establishing trust of published scientific content. This is true whether we are considering a data set, a computational workflow, a peer-reviewed publication or a simple scientific claim with supportive evidence. Existing vocabularies such as Dublin Core Terms (DC Terms) and the W3C Provenance Ontology (PROV-O) are domain-independent and general-purpose and they allow and encourage for extensions to cover more specific needs. In particular, to track authoring and versioning information of web resources, PROV-O provides a basic methodology but not any specific classes and properties for identifying or distinguishing between the various roles assumed by agents manipulating digital artifacts, such as author, contributor and curator.

**Results:** We present the Provenance, Authoring and Versioning ontology (PAV, namespace http://purl.org/pav/): a lightweight ontology for capturing "just enough" descriptions essential for tracking the provenance, authoring and versioning of web resources. We argue that such descriptions are essential for digital scientific content. PAV distinguishes between contributors, authors and curators of content and creators of representations in addition to the provenance of originating resources that have been accessed, transformed and consumed. We explore five projects (and communities) that have adopted PAV illustrating their usage through concrete examples. Moreover, we present mappings that show how PAV extends the W3C PROV-O ontology to support broader interoperability.

**Method:** The initial design of the PAV ontology was driven by requirements from the AlzSWAN project with further requirements incorporated later from other projects detailed in this paper. The authors strived to keep PAV lightweight and compact by including only those terms that have demonstrated to be pragmatically useful in existing applications, and by recommending terms from existing ontologies when plausible.

**Discussion:** We analyze and compare PAV with related approaches, namely Provenance Vocabulary (PRV), DC Terms and BIBFRAME. We identify similarities and analyze differences between those vocabularies and PAV, outlining strengths and weaknesses of our proposed model. We specify SKOS mappings that align PAV with DC Terms. We conclude the paper with general remarks on the applicability of PAV.

## Background

Research in the life sciences is becoming increasingly digital and collaborative. Scientists tend to conduct their investigations and reporting using digital resources (e.g., data artifacts, articles, etc.) obtained by aggregating existing resources (potentially generated as a result of other research investigations conducted by other scientists), and processed and analyzed using manual or automated workflows.

In such a context, scientists require a systematic means to organize and annotate resources [1]. This might require them, amongst other things, to: (i) trace the origin of a given resource; (ii) specify its previous and subsequent versions; and (iii) identify the creators (be they humans or machines) responsible for the existence of the resource, as well as the contributors who enriched and updated its content. For the most general use, a provenance vocabulary that meets these criteria must also be relatively compact and terse.

* Correspondence: paolo.ciccarese@gmail.com
†Equal contributors
[1]Department of Neurology, Massachusetts General Hospital, 55 Fruit Street, Boston, MA 02114, USA
[2]Harvard Medical School, 25 Shattuck Street, Boston, MA 02115, USA
Full list of author information is available at the end of the article

Several provenance vocabularies including Dublin Core Terms (DC Terms) [2], PROV-O [3], OPM [4], and Provenance Vocabulary [5] partially address these general needs, at varying levels of richness, complexity and maturity. In this section, after discussing the original use case, we discuss the strengths and weaknesses of these vocabularies and gaps in usage coverage, which led us to develop the Provenance Authoring and Versioning (PAV) ontology. In the Results section we also provide mappings from PAV to PROV-O and DC Terms.

### Original use case: SWAN platform

The Semantic Web Applications in Neuromedicine (SWAN) web-based collaborative platform [6] is an example of an application that embodies many of the above requirements. SWAN aims to organize and annotate scientific knowledge regarding neurodegenerative disorders and to facilitate the formation, development and testing of hypotheses. In particular, the AlzSWAN [7] knowledge base (AlzSWAN KB), a collaboration of SWAN's developers with the AlzForum web community of Alzheimer Disease researchers [8], is an instance of SWAN configured to allow the scientific community of Alzheimer Disease (AD) researchers to author, curate and connect a diversity of data and ideas about AD. The AlzSWAN curators typically read carefully a scientific article, usually representing a hypothesis on AD, and produce a linear representation of the embedded scientific discourse: claims, hypotheses and questions. For each of the discourse elements the curator selects related publications, proteins and genes. Knowledge in the AlzSWAN KB is shared using the SWAN Ontology [9] for interoperability.

One of the goals of the AlzSWAN KB consists in clearly recording the provenance of the digital artifacts as well as the provenance of the content or knowledge elements represented by the artifacts, and the agents (organizations, people and software) involved in creating and manipulating those artifacts. There is a clear distinction between the roles of the authors and curators, and the source of content:

**Authors** are the primary originators of scientific statements, originally conceiving the content (e.g. a tabular dataset).

**Curators** collect the knowledge published by the authors, interpreting and transforming the content of a textual document into SWAN research statements (hypothesis, claim or research question). They restructure the previously authored content and shape it to be appropriate for the intended representation (e.g. by normalizing the fields for being represented in a spreadsheet). Curators create the SWAN KB version that embodies the authors' work; thus they are contributing to the knowledge representation. However, the main intellectual property remains attributed to the original authors.

**Artifact creators** take care of physically creating the digital artifact by entering the statements and their links into the platform, (e.g. saves the spreadsheet as an *.xlsx* file).

**External sources** are the external data- and knowledge-bases such as PubMed [10] and UniProt [11,12] that AlzSWAN draws upon for metadata and for integrated data. Some of this metadata are retrieved and cached as they are, while some are imported after one or more transformations. It is important to track the original source and how it was incorporated in the knowledge base.

As depicted in Figure 1, the AlzSWAN knowledge capture and curation process consists of several steps:

A PhD-level neuroscientist (the *curator*) reads carefully an article (written by *authors*) usually representing a hypothesis on AD.

Based on the reading, the curator produces a textual document with a linear knowledge representation of the scientific discourse of the article, by building an ordered list of claims, hypotheses and questions.

For each of those elements, the curator identifies external resources such as related publications, proteins and genes. These resources provide data that can be retrieved in unmodified form, or imported after a transformation.

When possible, the formed representation is shared with the authors of the original article for collecting feedback.

The knowledge map is entered in AlzSWAN by a second person (the artifact *creator*) through a web user interface, which eventually encodes the textual content according to the SWAN ontology.

While step 5 represents the straightforward creation of the digital artifact, steps 1–4 represent the curation of the knowledge that the authors expressed in the journal article. Curation involves high-level domain knowledge and acts of judgement and creative composition. Both authors and curators originate digital content; *authorship* denotes the role of creative invention of a work, while the artifact *creator* of a work in our terminology is responsible for accurate transcription and encoding into final digital form.

Figure 1 also depicts tasks of revision, publishing and feedback collection. In particular, the feedback might motivate the generation of a new version of the encoded knowledge. Normally, in the case of AlzSWAN a new version might include newly available evidence supporting a given claim.

### Existing provenance vocabularies

There are existing vocabularies that at first appear to be promising to address the needs of AlzSWAN and similar

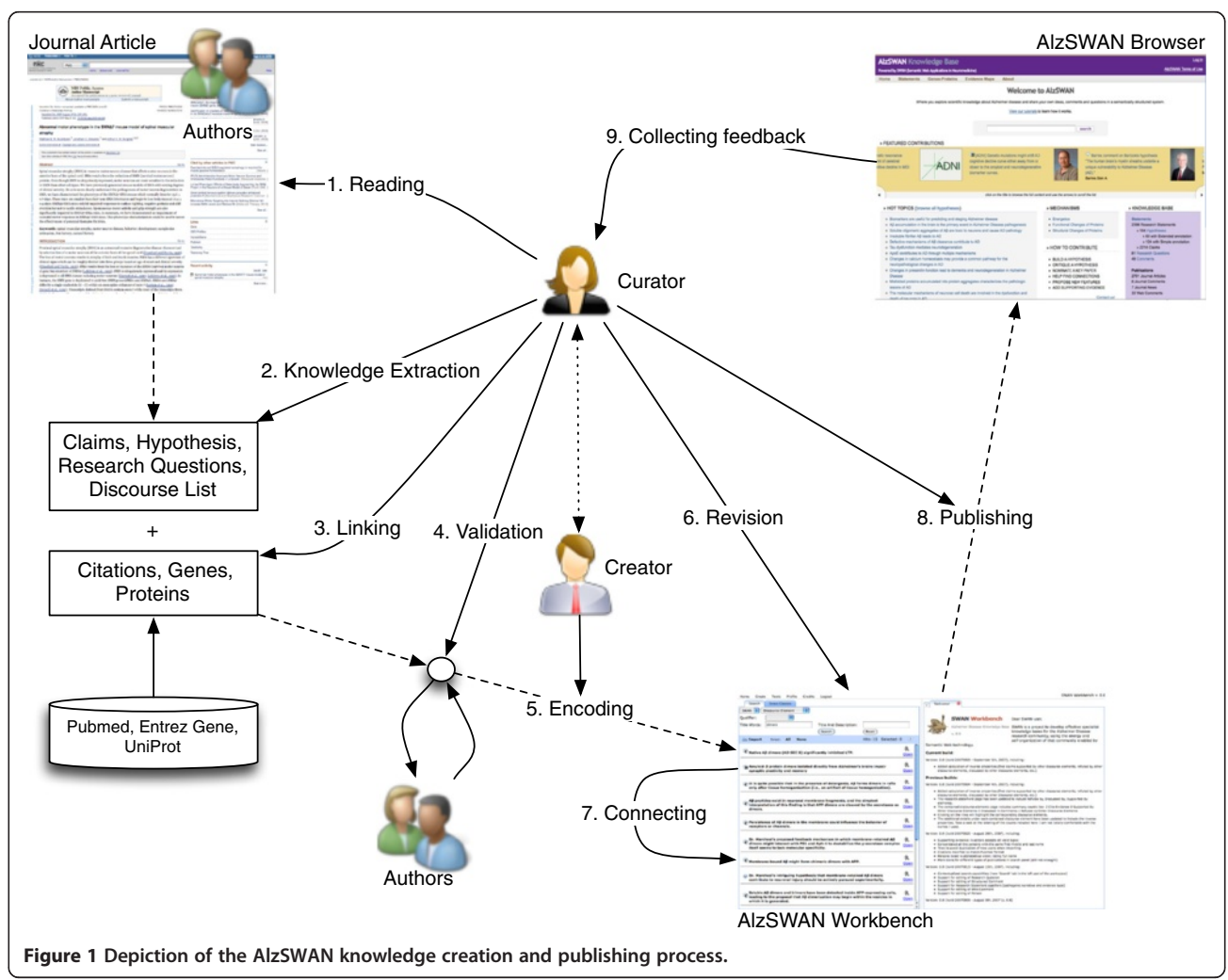

**Figure 1** Depiction of the AlzSWAN knowledge creation and publishing process.

applications: Dublin Core Terms (DC Terms) [2], PROV-O [3], OPM [4], and Provenance Vocabulary [5].

The Dublin Core Metadata Initiative (DCMI) provides core metadata vocabularies to support interoperable solutions for discovering and managing resources. In particular, DC Terms provides terms for specifying the entities that create and/or contribute to the existence of a given resource. While terms such as *dct:contributor* and *dct:creator* are useful and popular, we argue that they conflate what we consider as the distinct roles of contributor/author/curator and creator of the representation. For instance, a person who converts a web page from HTML 3 to HTML 5 could be said to be the *dct:creator* or *dct:contributor* of the new document, even when that person has not modified the human readable content of the document.

DC Terms provides a means for specifying derivation and representational differences (*dct:source, dct:isVersionOf, dct:isFormatOf*), but these do not clearly distinguish between a resource which was simply copied, one which was transformed to give rise to a new resource, or one which was further derived by adding additional content (for example new scientific evidence). For versioning DC Terms has *dct:isVersionOf*, which implies *substantive changes in content* (e.g. a movie can be a version of a theater play), and *dct:replaces*, which indicates a superseded resource. However, these statements do not quite fit with dealing with and distinguishing between: smaller updates (e.g. spelling mistakes); larger derivations (which might no longer be versions of the original); and linear revision history without necessarily indicating previous versions as outdated. Consider, for example, a health authority where Dr. Doe authored and implemented the clinical guidelines for treating hypertensive patients and sent it to Dr. Green for comments. Green edits his copy to suggest extending a drug treatment from 2 weeks to 4 weeks. Because Green's guideline has not been approved yet, we cannot state that his document replaces (*dct:replaces*) Doe's document. Also, Green's document is in the same form and does not introduce substantive changes (*dct*: *isVersionOf*) from Doe's document.

PROV-O [3], an OWL ontology developed by the W3C Provenance Working group [13] aims to provide a standard for representing and exchanging domain-independent provenance information between applications and systems. PROV-O provides terms that can be used to trace the origin of a given resource, its derivation history, as well as the relationship between the resources, and the entities that contributed to the existence of the resource. PROV-O can be used at detailed process level with *activity-agent-entity* interactions, or at a higher level with shortcuts for *entity-entity* and *entity-agent* relations such as *prov:wasDerivedFrom* and *prov:wasAttributedTo*.

Overall PROV-O is a generic provenance data model, which can be extended to give domain-specific provenance, for instance by subclasses of *prov:Activity* or sub-properties of *prov:wasAttributedTo*. PROV-O does not itself provide any distinctions between authors, curators, contributors or artifact creators.

PROV-O specifies a set of common extensions, which at first glance would seem to cover some of our requirements: i) *prov:hadPrimarySource*: a kind of "derivation relation from secondary materials to their primary sources, which were produced by some agent with direct experience and knowledge about the topic", ii) *prov:wasQuotedFrom*, the "repeat of (some or all of) an entity, such as text or image, by someone who may or may not be its original author".

However, on closer examination these terms are inadequate. For example, in the AlzSWAN project, curators consulted papers on the web. Such an action cannot be described using *prov:hadPrimarySource*, because the documents consulted by the curators were not necessarily *primary sources* [3] such as witness statements, reports or interviews; we found the definition of primary sources to be too narrow to cover most academic publications. Instead, we need a means to describe the fact that the curators simply accessed a document. Curators can also download a file from a source on the web, such as UniProt. Although *prov:wasQuotedFrom* can be used to relate the document to the source it was downloaded from, it does not reflect that the document is a complete and exact copy of the document in the source.

The Open Provenance Model [4] predates PROV-O, and has a very similar approach to modeling provenance by relating agents, artifacts and processes. By and large, the concepts of OPM are covered by equivalent PROV-O concepts, therefore the above analysis of PROV-O applies also to OPM.

DC Terms, PROV-O and OPM are domain-independent and general-purpose vocabularies. Therefore the limitations we identified above should not be perceived as issues that need to be addressed within these vocabularies. In fact, the PROV-O authors do not claim that the vocabulary is complete. Instead, they encourage users to extend it with terms that capture their domain needs.

The Provenance Vocabulary [5] describes data access, creation, retrieval and publishing as detailed chains of *prov*: *Activity* and *prov:Entity* relations. Terms like *prv:accessedResource*, *prv:createdBy*, *prv:retrievedBy* seem relevant for our use cases, but as detailed in the Discussion section, this approach come at the cost of increased verbosity, which we argue reduces the ability to query the provenance in a consistent way.

For describing the provenance of the AlzSWAN use cases such as the one depicted in Figure 1, we designed the *PAV* (*Provenance, Authoring and Versioning*) ontology, whose most recent version, PAV 2, maps to PROV-O. In PAV, we do not attempt to model the whole chain of process-oriented provenance like Open Provenance Model (OPM) processes [4] and PROV-O activities [3], or the many forms of metadata as covered by Dublin Core Terms (DC Terms) [2]. Rather, in PAV we focus on the provenance of a digital resource in terms of its relationships with other digital resources and agents involved in their creation, authoring and manipulation, and we abstract away from the description of the activities (process) that manipulate and transform the digital resources.

## Results

In this section we present the PAV ontology, describing its structure and constituent terms. We go on to present the systems and communities that have adopted PAV, and to discuss how they use it. Finally, we present a collection of mappings, specifying how PAV extends the W3C PROV-O vocabulary.

### PAV Ontology

PAV is a lightweight vocabulary, for capturing "just enough" descriptions essential for web resources representing digitized knowledge. PAV is intended to specify *Provenance*, *Authoring* and *Versioning* information. Accordingly, this section is organized into three subsections. PAV properties are outlined in **bold** where they are defined, described or exemplified; and in *italics* elsewhere. Properties belonging to other vocabularies are always in italics.

The *pav*: prefix indicates the namespace http://purl.org/pav/ [14], which also resolves to the latest version of the PAV ontology as OWL. PAV is currently in version 2.2 (OWL importable from http://purl.org/pav/2.2 [15]). Further details of versions and changes are listed on the wiki [16].

**Authoring**

In scholarly communication, it is crucial to be able to precisely attribute the several forms of authorship (intellectual property) or contributions of the knowledge content and of its representation [1]. The PAV ontology provides properties for tracking intellectual property information, which are described in Table 1.

As suggested by the properties listed in Table 1, in PAV we distinguish between authors that originate or creatively invent a work that is expressed in a digital resource (**pav:authoredBy**), e.g., the authors of a scientific publication or of a novel scientific hypothesis; and curators (**pav:curatedBy**), who are content specialists responsible for shaping the expression in an appropriate format. When talking about knowledge artifacts the authors are contributing the primary knowledge and the curators are those responsible for updating the knowledge base. Contributors (identified by the superproperty **pav:contributedBy**) cover both authors and curators, as well as agents that generically provide some help in conceiving the resource or in the expressed knowledge creation/extraction. For example, a scientist who performed some biological experiments and published the results in a paper is considered an author (*pav:authoredBy*) as she produced novel results. The agent that analyzes the paper, extracts and organizes some of the scientific discourse in argumentation – hypotheses, claims, etc. – is the curator (*pav:curatedBy*). Finally the person that enters such knowledge in a hypotheses management application is the creator of the knowledge artifacts (*pav:createdBy*), as defined in the next subsection.

As illustrated in Table 1, PAV authoring properties can be associated with a timestamp using the following properties: **pav:authoredOn**, **pav:curatedOn**, and **pav:contributedOn**.

For describing the publication process of the created resource, we recommend adopting the following DC Terms properties: *dct:publisher*, *dct:issued*, *dct:dateSubmitted*, *dct:dateAccepted* and *dct:dateCopyrighted*. For instance, *dct:publisher* identifies 'an entity responsible for making the resource available' [2]. It is important to note that these properties describe the publication of the particular resource (say a knowledge graph), not the publication of the original work that this resource might have been imported or derived from. The PAV term *pav:*

**Table 1 PAV authoring properties**

| | |
|---|---|
| pav:authoredBy | Indicates an agent that originated or gave existence to the work that is expressed by the digital resource. The author of the content of a resource may be different from the creator of that resource representation (*pav:createdBy*), although they are often the same. |
| | *pav:authoredBy* is more specific than its superproperty *dct:creator* - which might or might not be interpreted to also cover the creation of the representation of the artifact. |
| | The author is usually not a software agent (which would be indicated with *pav:createdWith*, *pav:createdBy* or *pav:importedBy*), unless the software actually authored the content itself; for instance an artificial intelligence algorithm which authored a piece of music or a machine learning algorithm that authored a classification of a tumor sample [17]. |
| pav:authoredOn | Indicates the date this resource was authored by the agents given by *pav:authoredBy*. Note that *pav:authoredOn* is different from *pav:createdOn*, although their values are often the same. |
| | This property is normally used in a functional way, indicating the last time of authoring, although PAV does not formally restrict this. |
| pav:curatedBy | Specifies an agent specialist responsible for shaping the expression in an appropriate format. Often the primary agent responsible for ensuring the quality of the representation. The curator may be different from the author (*pav:authoredBy*) and creator of the digital resource (*pav:createdBy*). The curator may in some cases be a software agent, for instance text mining software which adds hyperlinks for recognized genome names. |
| pav:curatedOn | Specifies the date this resource was curated. *pav:curatedBy* gives the agents that performed the curation. |
| | This property is normally used in a functional way, indicating the last curation date, although PAV does not formally restrict this. |
| pav:contributedBy | Specifies an agent that provided any sort of help in conceiving the work that is expressed by the digital artifact. |
| | Contributions can take many forms, of which PAV define the subproperties *pav:authoredBy* and *pav:curatedBy*; however other specific roles could also be specified by *pav:contributedBy* or custom subproperties, such as illustrating, investigating or managing the underlying data source. Contributions can additionally be expressed in detail using *prov:qualifiedAttribution* and *prov:hadRole*. |
| | Note that *pav:contributedBy* identifies only agents that contributed to the work, knowledge or intellectual property, and not agents that made the digital artifact or representation (*pav:createdBy*), thus the considerations for software agents is similar to for *pav:authoredBy* and *pav:curatedBy* above. |
| | *pav:contributedBy* is more specific than its superproperty *dct:contributor* - which might or might not be interpreted to also cover contributions to making the representation of the artifact. |
| pav:contributedOn | Indicates the date this resource was contributed on. *pav:contributedBy* specifies the agents that contributed. This term is a superproperty of *pav:authoredOn* and *pav:curatedOn*, but can also be used for other kinds of contributions, such as illustrating or investigating. |

**providedBy** (see next subsection) gives a shortcut to express the publisher of the original work.

Figure 2 is an example illustrating the usage of PAV authoring terms. The figure depicts a partial representation of a hypothesis taken from the AlzSWAN knowledge base and published by the AlzSWAN team on behalf of AlzForum. Such a claim has been derived from a scientific publication and therefore is recorded as authored by the publication author. A curator performed on a particular date the task of encoding the research statements into the SWAN format.

### Provenance

To encode provenance information specifying creation, retrieval, import and source access, PAV provides the properties presented in Table 2.

While authoring terms like *pav:authoredBy* and *pav:contributedBy* describe who brought the underlying knowledge to light, the provenance term **pav:createdBy** describes who created the digital resource. For instance, in Figure 2, the creator is the user who formally encodes the claim, while the author is the person who wrote the original published article. In PAV, the digital artifact was **pav:createdOn** a given date, and it was **pav:createdWith** a specific software application. Similarly, a digitization of Charles Darwin's Galápagos notebook could be *pav:createdOn* 2006-10-06 T09:49:12Z and *pav:authoredOn* 1835-03-06 T00:00:00Z. (Note that this *xsd:dateTime* string uses the convention of a zero timestamp as the exact time of day is unknown).

In PAV, we also distinguish between retrieving a resource 'as is' (**pav:retrievedFrom**), such as caching or downloading; importing a resource through a data transformation (**pav:importedFrom**) in order to fit it into an existing model, e.g., when converting a CSV file to an Excel spreadsheet; and accessing a resource (**pav:sourceAccessedAt**). The latter is useful when resources such as webpages are accessed but not cached or imported into the system.

As well as the above properties, PAV allows us to specify the agent that performed the task – **pav:retrievedBy**, **pav:importedBy**, **pav:sourceAccessedBy** – and the time when the task was performed – **pav:retrievedOn**, **pav:importedOn** and **pav:sourceAccessedOn**. For example, Figure 3 is a snippet specifying the record of a protein generated by importing data from the EntrezGene database [18]. The relationship *pav:importedFrom* is used to assert that the record is the result of a transformation process. In this specific case, an XML file has been translated into RDF according to a specific model (for example the *lses* namespace from the SWAN Ontology). The transformation is attributed to the agent that performed it – in this particular case a software agent – through the relationship *pav:importedBy*.

Additional properties can be used to enrich the provenance data. For instance **pav:createdAt** provides the geolocation of the agent when the artifact has been created.

### Versioning and evolution

To avoid complexity, PAV adopts a 'snapshot'-based approach as opposed to a detailed process-oriented approach.

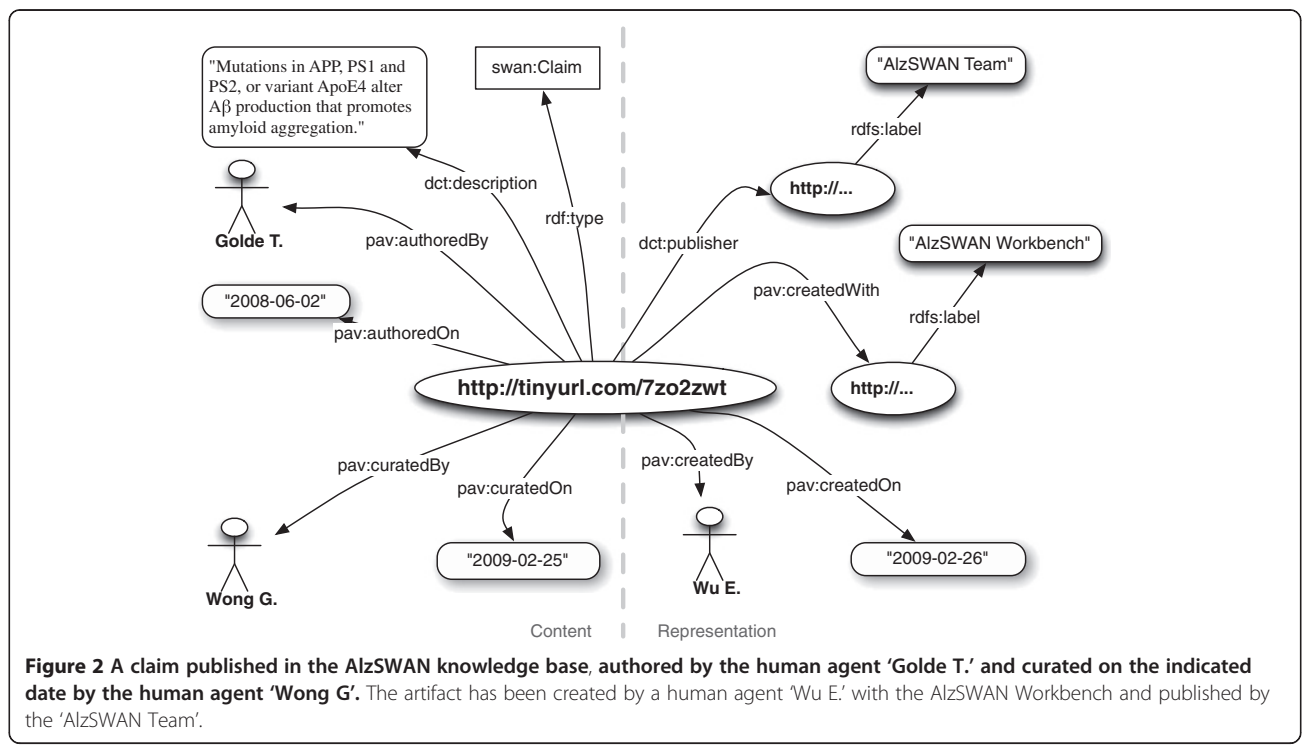

**Figure 2 A claim published in the AlzSWAN knowledge base**, **authored by the human agent 'Golde T.' and curated on the indicated date by the human agent 'Wong G'.** The artifact has been created by a human agent 'Wu E.' with the AlzSWAN Workbench and published by the 'AlzSWAN Team'.

**Table 2 PAV provenance properties**

| | |
|---|---|
| pav:createdBy | An agent primarily responsible for encoding the digital artifact or resource representation. This creation is distinct from forming the content, which is indicated with *pav:contributedBy* or its subproperties. |
| | *pav:createdBy* is more specific than its superproperty *dct:creator* - which might or might not be interpreted to also cover the creation of the content of the artifact. |
| | For instance, the author wrote' this species has bigger wings than normal' in his log book. The curator, going through the log book and identifying important knowledge, formalizes this as 'locus perculus has wingspan > 0.5 m'. The artifact creator enters this knowledge as a digital resource in the knowledge system, thus creating the digital artifact (say as JSON, RDF, XML or HTML). |
| | A different example is a news article. *pav:authoredBy* indicates the journalist who wrote the article. *pav:contributedBy* can indicate the artist who added an illustration. *pav:curatedBy* can indicate the editor who made the article conform to the news paper's language style. *pav:createdBy* can indicate who put the article on the web site. |
| | The software tool used by the creator to make the digital resource (say Protege, Wordpress or OpenOffice) can be indicated with *pav:createdWith*. |
| pav:createdOn | The date of creation of the digital artifact or resource representation. The agents responsible can be indicated with *pav:createdBy*. |
| | This property is normally used in a functional way, indicating the time of creation, although PAV does not formally restrict this. *pav:lastUpdateOn* can be used to indicate minor updates that did not affect the creating date. |
| pav:createdWith | The software/tool used by the creator (*pav:createdBy*) when making the digital resource, for instance a word processor or an annotation tool. A more independent software agent that creates the resource without direct interactions by a human creator should instead be indicated using *pav:createdBy*. |
| pav:createdAt | The geo-location of the agents when creating the resource (*pav:createdBy*). For instance, a photographer takes a picture of the Eiffel Tower while standing in front of it. |
| pav:retrievedFrom | The URI where a resource has been retrieved from. Retrieval indicates that this resource has the same representation as the original resource. If the resource has been somewhat transformed, *pav:importedFrom* should be used instead. This property is normally used in a functional way, although PAV does not formally restrict this. |
| pav:retrievedBy | An entity responsible for retrieving the data from an external source. The retrieving agent is usually a software entity, which has done the retrieval from the original source without performing any transcription. |
| | Retrieval indicates that this resource has the same representation as the original resource. If the resource has been somewhat transformed, use *pav:importedFrom* instead. |
| pav:retrievedOn | The date the source for this resource was retrieved. This property is normally used in a functional way, although PAV does not formally restrict this. |
| pav:importedFrom | The original source of imported information. Import means that the content has been preserved, but transcribed somehow, for instance to fit a different representation model by converting formats. The imported resource does not have to be complete but should be consistent with the knowledge conveyed by the original resource. |
| pav:importedBy | An agent responsible for importing data from a source given by *pav:importedFrom*. The importer is usually a software agent which has done the transcription from the original source. Note that *pav:importedBy* may overlap with *pav:createdWith*. |
| pav:importedOn | The date the resource was imported from a source given by *pav:importedFrom*. This property is normally used in a functional way, indicating the first import date, although PAV does not formally restrict this. |
| | This property is normally used in a functional way, although PAV does not formally restrict this. If the resource is later reimported, this should instead be indicated with *pav:lastRefreshedOn*. |
| pav:lastRefreshedOn | The date of the last import of the resource. This property is used if this version has been updated due to a re-import, rather than the import creating new resources related using *pav:previousVersion*. |
| pav:providedBy | The *original provider* of the encoded information (e.g. PubMed, UniProt, Science Commons). |
| | The provider might not coincide with the *dct:publisher*, which would describe the *current* publisher of the resource. For instance if the resource was retrieved, imported or derived from a source, that source was published by the original provider. *pav:providedBy* provides a shortcut to indicate that original provider on the new resource. |
| pav:sourceAccessedAt | A source which was accessed or consulted (but not retrieved, imported or derived from). For instance, a curator (*pav:curatedBy*) might have consulted figures in a published paper to confirm that a dataset was correctly *pav:importedFrom* the paper's supplementary CSV file. |
| | Another example: I can access the page for tomorrow weather in Boston (http://www.weather.com/weather/tomorrow/Boston+MA+02143) and I can blog 'tomorrow is going to be nice'. The source does not make any claims about the nice weather, that is my interpretation; therefore the blog post has *pav:sourceAccessedAt* the weather page. |
| pav:sourceAccessedBy | The agent who accessed the source given by *pav:sourceAcccessedAt* . |

**Table 2 PAV provenance properties** *(Continued)*

| | |
|---|---|
| pav:sourceAccessedOn | The date when the original source given by *pav:sourceAccessedAt* was accessed to create the resource. |
| | For instance, if the source accessed described the weather forecast for the next day, the time of source access can be crucial information. |
| | This property is normally used in a functional way, although PAV does not formally restrict this. If the source is subsequently checked again (say to verify validity), this should be indicated with *pav:sourceLastAccessedOn*. |
| pav:sourceLastAccessedOn | The date when the original source given by *pav:sourceAccessedAt* was last accessed and verified, especially when the source has previously been *pav:sourceAccessedOn* when creating the resource. This property is normally used in a functional way, although PAV does not formally restrict this. |
| | This property can be useful together with *pav:lastRefreshedOn* or *pav:lastUpdateOn*, but could also be used alone, for instance when a source was verified and no further action was taken for the resource. |

Snapshots are identified by a URI and the free text property **pav:version**, and several snapshots of the same resource are related using **pav:previousVersion**, linking a version of the resource with the previous one of the same lineage. We use **pav:derivedFrom** to indicate an artifact as a derivation of another, not necessarily of the same lineage. Table 3 presents and describes PAV versioning properties.

As an illustrative example, Figure 4 specifies that the Hypothesis A1 is first created and then updated into A1' by the same human agent. A1 and A1' are two representation of the same version of the same hypothesis at two different points in time. The version A2 of the same hypothesis is created by another human agent having the same access rights. From the digital artifact standpoint it is accurate to say that the second version (A2) was created by a different agent than the first one, even if both versions are of the same lineage. In PAV the relationship *pav:createdBy* does not have any content authorship connotation. Computing the differences between the two versions allows attributing the intellectual property to the authors/curators who originated it. It is also possible to branch the lineage of digital artifacts through the relationship *pav:derivedFrom*. The property **pav:lastUpdateOn** is used to date when the digital artifact was last updated, indicating minor changes that did not signify a change of version (and therefore a new resource), such as fixing a typographical error.

Dublin Core Terms provides the property *dct:source*, which is used to specify "A Reference to a resource from which the present resource is derived. The present resource may be derived from the Source resource in whole or part". Note, however, that this is more permissive, and therefore less specific, than *pav:derivedFrom* as it encompasses also format conversions.

**Multiplicity**

The properties defined by PAV do not have any multiplicity constraints. That means it is valid, for instance, to specify multiple authors using *pav:authoredBy*, or multiple contribution dates using *pav:contributedOn*. Some of the properties, like *pav:retrievedFrom*, *pav:lastRefreshedOn* or *pav:version* should still primarily be used in a functional way (as indicated in Tables 1, 2 and 3), as their interpretation could be difficult with multiple values. However PAV does not formally add functionality constraints to the OWL ontology.

Combinations of multiple agents, sources and dates may be used with PAV, but would mean that finer details are not fully preserved, such as *who* accessed *which* source *when*. This is the result of a compromise between simplicity and completeness when designing PAV. It is recommended that applications that want to keep those details also provide an accompanying PROV-O trace, as exemplified in Figure 5.

In this example PAV will tell us there are 3 authors, and but the resource has only got a single authoring date (the last time of authoring). We do not know from PAV alone when the different authors participated, but the expanded PROV-O trace, which qualifies the implied *prov:wasAttributedTo* relations, can detail those dates using *prov:atTime*. Here the nature of the individual attributions are also indicated using *prov:hadRole* and a custom role vocabulary *ex:*, and the *prov:atLocation* indicates that *:khalid* and *:stian* were in the same office, authoring at the same time.

In some cases it might not be appropriate to attribute the individual agents directly, for instance a report authored by a committee where the individual members have discussed and voted over the content. We recommend for such cases to be represented as a single identified agent, typically a *prov:Organization*, and the individual members represented using the Collection Ontology [20], as shown in Figure 6.

```
_:geneRecord a lses:GeneRecord;
      lses:fullName "Amyloid beta (A4) precursor protein" ;
      lses:preferredSymbol "APP" ;
      pav:importedOn "2009-02-26T19:49:12-0500"^^xsd:dateTime ;
      pav:importedFrom <http://www.ncbi.nlm.nih.gov/> ;
      pav:importedBy [
        a agent:Software ;
        rdfs:label "AlzSWAN" ;
          agent:softwareVersion "2.0" ] .
```
**Figure 3 Example of import from the EntrezGene database expressed using Turtle notation [19].** The two namespaces *lses* (life science entities [http://purl.org/swan/1.2/lses/] and *agents* [http://purl.org/swan/1.2/agents/] are part of the SWAN suite of ontologies.

**Table 3 PAV versioning properties**

| | |
|---|---|
| pav:version | The version identifier of a resource. This is a free text string, typical values are '1.5' or '21'. The URI identifying the previous version can be provided using *pav:previousVersion*. This property is normally used in a functional way, although PAV does not formally restrict this. |
| pav:previousVersion | The previous version of a resource in a lineage. For instance a news article updated to correct factual information would point to the previous version of the article with *pav:previousVersion*. If, however, the content has significantly changed so that the two resources no longer share lineage (say a new article that talks about the same facts) they can instead be related using *pav:derivedFrom*. |
| | This property is normally used in a functional way, although PAV does not formally restrict this. A version identifier for a resource can be provided using the data property *pav:version*. |
| pav:derivedFrom | Derived from a different resource. Derivation concerns itself with derived knowledge. If this resource has the same content as the other resource, but has simply been transcribed to fit a different model (like XML to RDF or SQL to CSV), use *pav:importedFrom*. If the content has been further refined or modified, use *pav:derivedFrom*. |
| | Details about who performed the derivation (e.g. who did the refining or modifications) may be indicated with *pav:contributedBy* and its subproperties. |
| pav:lastUpdateOn | The date of the last update of the resource. An update is a change which did not warrant making a new resource related using *pav:previousVersion*, for instance correcting a spelling mistake. This property is normally used in a functional way, although PAV does not formally restrict this. |

### Contribution roles

While PAV properties like *pav:retrievedBy* and *pav:createdWith* are fairly specific, yet generally applicable; different domains will vary in their understanding of what in PAV would constitute the roles of *pav:authoredBy*, *pav:curatedBy* or *pav:contributedBy*. For instance, the International Workshop on Contributorship and Scholarly Attribution 2012 [21] presented a survey where "authorship" was found to regularly include roles such as design of experimental methods and statistical analysis, but also software development, preparing graphics and managing a laboratory. The presented text analysis on acknowledgement sections in academic papers identified non-author contributions like funding, technical assistance, data contribution, and animal assistance.

It is out of scope for PAV to try to model this wide range of contributorships, but we do note that ontologies such as SPAR's Publishing Roles Ontology [22] define roles like *pro:illustrator*, *pro:critic* and *pro:editor* which would be appropriate to use with *prov:hadRole* in the pattern shown in Figure 5. Additionally, third-party subproperties, e.g. of *pav:contributedBy* and *pav:authoredBy*, can be created to further specify the form of the contribution, utilizing PAV as a common platform for attributions across domains.

### Who is using PAV

PAV has successfully been applied by several projects in academia and in industry due to it being compact and easy to understand. Besides the SWAN project, for which PAV was originally developed, it has been used in the Annotation Ontology (AO) [23]; the Domeo Annotation Tool [24]; the Nanopublications specification [25]; the Open PHACTS dataset description specification [26-28]; the Wf4Ever Research Objects [29]; and the Elsevier Satellite article annotation format [30]. For its most recent release, we updated PAV to include a mapping to the PROV-O ontology [3],

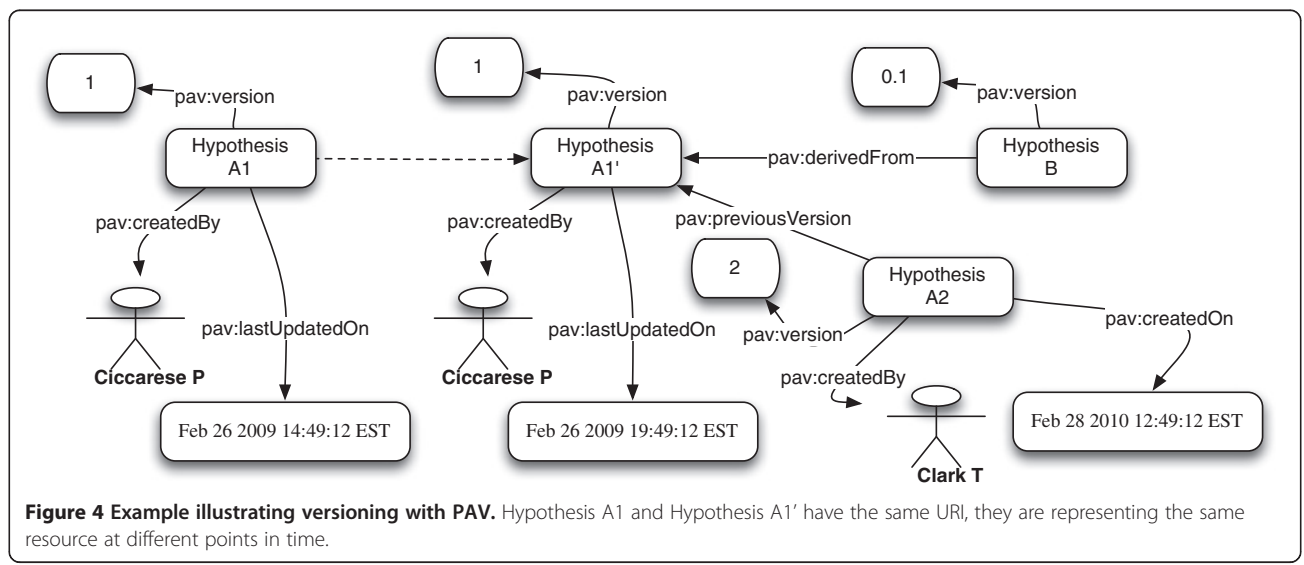

**Figure 4 Example illustrating versioning with PAV.** Hypothesis A1 and Hypothesis A1' have the same URI, they are representing the same resource at different points in time.

```
<http://example.com/resource.html>
    pav:authoredBy :paolo, :stian, :khalid ;
    pav:authoredOn "2013-02-20T15:19:10+05:00"^^xsd:dateTime ;
    prov:qualifiedAttribution
      [ prov:agent :paolo ;
        prov:hadRole ex:editor ;
        prov:atTime "2013-02-20T15:19:10+05:00"^^xsd:dateTime ] ,
      [ prov:agent :stian ;
        prov:hadRole ex:coder ;
        prov:atTime "2013-02-18T10:21:54+00:00"^^xsd:dateTime ;
        prov:atLocation :manchesterOffice ] ,
      [ prov:agent :khalid ;
        prov:hadRole ex:explainer ;
        prov:atTime "2013-02-18T10:21:54+00:00"^^xsd:dateTime ;
        prov:atLocation :manchesterOffice ] .
```

**Figure 5 Example illustrating how PROV-O can be combined with PAV in order to provide a more detailed provenance record.**

enabling a level of compatibility for all PROV-based tools for analyzing and combining provenance. Figure 7 illustrates PAV's relation to these projects.

### Annotation ontology and Domeo annotation tool

PAV is used extensively by many of the applications making use of the Annotation Ontology (AO) [23] for anchoring annotations to online resources. One such application is the Domeo Web Annotation Toolkit [24], a collection of software components that provides a rich set of features including:

- semantically annotating online HTML and XML documents;
- sharing the annotation in RDF; and
- searching the annotation while leveraging semantic inference.

Domeo, as well as the other AO applications, provide a constant stream of requirements and feedback for testing and improving the PAV model.

In Figure 8 we represent a common scenario where the annotation artifact is digitizing an annotation that has been originally performed on the physical manifestation of a picture. In other words: Khalid scribbled a note on a classic printed picture; Paolo found a digital version of that picture and interpreted the handwritten note by Khalid and passed it along to Stian who, using Domeo, created an AO artifact representing the whole scenario. Khalid is the author (*pav: authoredBy*) of the original note, Paolo is the curator (*pav:*
*curatedBy*) of that content and Stian is the creator (*pav:createdBy*) of the digital artifact.

### Open PHACTS dataset descriptions

The aim of the Open PHACTS project [26,27] is to facilitate improvements in drug discovery using semantic web standards and technologies. Drug discovery requires the integration of data from many data sources covering chemical compounds (e.g. ChemSpider [31] and ChEMBL [32]), proteins (e.g. UniProt [11,12]), and drug interactions (e.g. DrugBank [33]). As such, the project requires accurate descriptions of the datasets they have used, identifying the particular versions, so that data provenance can be returned to the users.

In the specification [28], Open PHACTS recommend the use of existing vocabularies: Vocabulary of Interlinked Datasets (VoID) [34]; Dublin Core Terms (DC Terms) [2]; Friend of a Friend (FOAF) [35]; and PAV. VoID itself does not define any new provenance terms, but specifies a pattern of using DC Terms properties for purposes of recording provenance, such as *dct:creator*, *dct:contributor* and *dct:source* [36].

The Open PHACTS dataset specification, which uses a specialization of VoID, also specifies patterns of recording provenance, but from the provenance-related terms of DC Terms, only uses *dct:publisher* and *dct:issued*, and primarily recommends the PAV properties: *pav:version*, *pav: previousVersion*, *pav:retrievedFrom*, *pav:importedFrom*, *pav: importedOn*, *pav:importedBy*, *pav:derivedFrom*, *pav:createdOn*, *pav:createdBy*, *pav:createdWith*, *pav:authoredBy*, *pav: authoredOn*, *pav:lastRefreshedOn* and *pav:lastUpdateOn*.

The Open PHACTS VoID editor [37], shown in Figure 9, provides a wizard-like web interface for generating dataset descriptions, including the above-mentioned PAV properties.

### Nanopublications

Nanopublications [38] have been proposed by the ConceptWeb Alliance and the Open PHACTS project as a new means for publishing and citing specific core scientific statements. A nanopublication is composed of two basic elements: an assertion, and the provenance of that assertion. The former is used to express a single scientific fact, whereas the second is used to provide supporting evidence, in addition to the nanopublication attribution, i.e. its authors and other metadata information. To express provenance information, the Nanopublication specification [25] recommend the use of PAV and Open Provenance Model (OPM) [4]. For instances, the examples of nanopublications advertised on the

```
<http://example.com/report.html> pav:authoredBy <http://example.com/committee> ;
<http://example.com/committee> a prov:Organization, foaf:Organization, co:Set ;
    co:element :stian, :khalid, :paolo .
```

**Figure 6 Example illustrating how Collection Ontology (CO) can be used with PAV in order to encode a collection (set) of people.**

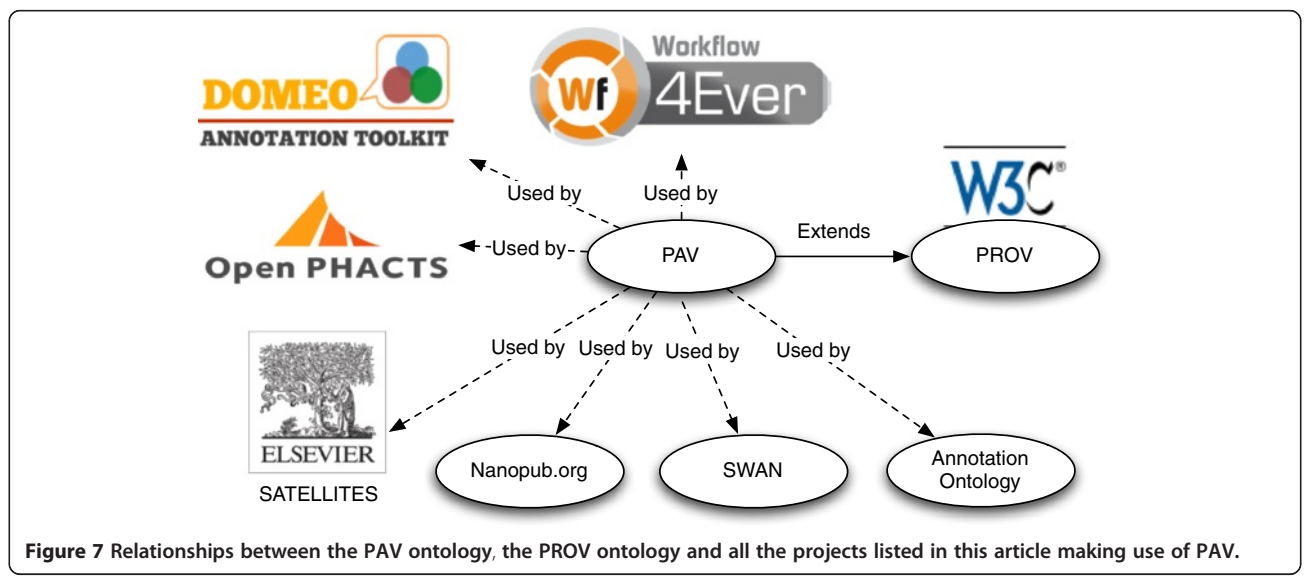

**Figure 7** Relationships between the PAV ontology, the PROV ontology and all the projects listed in this article making use of PAV.

nanopublication official website [39] make use of the following PAV properties: *pav:authoredBy*, *pav:createdBy*, *pav:version*.

To illustrate this use of PAV, the RDF snippets in Figures 10, 11 and 12 specify a nanopublication assertion and its provenance. These snippets are taken from a real world nanopublication [40], here slightly simplified for brevity. Notice that the attribution information (Figure 12) uses *pav:authoredBy* and *pav:createdBy* to distinguish the author (the scientist making the claim) from the artifact creator (who formed the nanopublication as RDF). The nanopublication itself is given a version number with *pav:version*.

Note that the process of structuring a nanopublication could also be considered a kind of curation, as in order to serialize the RDF the creator has to also look up identifiers for genes and symptoms, a content refinement as

```
:description a ao:Annotation ;
  ao:annotatesResource <http://example.com/picture.jpg> ;
  ao:body <http://khalid.example.org/aboutPicture.txt> ;
  pav:createdWith _:domeo ;
  pav:createdBy _:stian ;
  pav:createdOn "2012-12-14T15:02:14Z"^^xsd:dateTime ;
  pav:curatedBy _:paolo ;
  pav:curatedOn "2012-12-10T09:12:44Z"^^xsd:dateTime ;
  pav:importedBy _:bot .

<http://khalid.example.org/aboutPicture.txt>
  pav:authoredBy _:khalid ;
  dct:description "Text of the handwritten comment".
```

**Figure 8** Example of annotation representation using Annotation Ontology and PAV.

such identifiers presumably where not in the original content. The nanopublication we found as an example did however not specify *pav:curatedBy*, so from the above we can't tell anything more about the curation aspect of the nanopublication.

**Elsevier satellite**

PAV has been used, in conjunction with DC Terms, as the provenance ontology in the Elsevier Satellite annotation format [30]. The Satellite format is a linked data compliant data format to capture, store and expose metadata objects using open standards based metadata frameworks e.g. SKOS [43], DCMI and SWAN. Satellite format uses PAV 1.2 as originally specified in the SWAN suite of ontologies [9].

In Satellite format, PAV differentiates the provenance properties used for the metadata container and those used for the contained metadata items. Satellite uses the *dct:date* and *dct:creator* predicates to provide information on the item being described by the metadata. Satellite uses the *pav:createdOn* and *pav:createdBy* predicates in a header to describe the origin of the Satellite metadata itself.

**Research objects**

The notion of "Research Object" [29] was coined by the Wf4Ever project as an abstraction for the management of containers of sets of objects. In Wf4Ever, research objects bundle investigation-related resources such as documents, presentations, workflows, datasets, etc. Research Objects, their constituent resources, and their relationships, are described and annotated using existing vocabularies, including PAV. Specifically, the Wf4Ever Research Object Model [44] uses PAV to:

**Figure 9 The Open PHACTS VoID editor [37], a web-based wizard for creating a VoID dataset description, here representing a data format conversion by using PAV properties *pav:importedFrom*, *pav:importedOn*, *pav:importedBy*.** The VoID description itself (the generated RDF) has its own provenance, using *pav:createdBy*, *pav:createdOn*.

- Identify and distinguish the people ("Agents") who *contributed* to a Research Object (and its constituent resources) from those who *created* them. For instance, a hypothesis document could have been *authored* by a PhD student, but uploaded to a research object by their supervisor (thus *creating* the digital resource)
- State the time at which a Research Object, or a resource thereof, was *last updated*.
- Track the *versions* and origin of replicated resources, such as recording provenance for resources and user annotations which have been *imported from* and *retrieved from* third-party repositories using automated tools.

### PROV-O and PAV mapping

The earlier version 1.2 of PAV was recognized (as a component of the SWAN ontology) by the W3C Provenance Incubator Group [45] and was one of the foundational models [46] on which the requirements for the general provenance model later defined by the W3C Provenance Working Group.

```
{
    <#> a np:Nanopublication ;
        np:hasAssertion       :Assertion ;
        np:hasPublicationInfo :PublicationInfo ;
        np:hasProvenance      :Provenance .

    :Assertion       a nanopub:Assertion .
    :PublicationInfo a nanopub:PublicationInfo .
    :Provenance      a nanopub:Provenance .
}
:Provenance {
    :Assertion prov:wasDerivedFrom
                     textmining:gene_disease_concept_profiles_matching_1980_2010 ;
               prov:wasGeneratedBy textmining:gene_disease_concept_profiles_1980_2010 .
}
```

**Figure 10 Gene disease nanopublication example, in TriG format, *adapted from [40]*.** The nanopublication is expressed as three named graphs: the *Assertion* which expresses the claim of this nanopublication (Figure 11), the *PublicationInfo*, which details the attributions of this assertion (Figure 12), and *Provenance*, relating this nanopublication to the original data it was derived from (shown above).

```
:Assertion {
    :Association_1 a sio:statistical-association ;
        sio:has-measurement-value :Association_1_p_value ;
        sio:refers-to   <http://bio2rdf.org/geneid:55835>, <http://bio2rdf.org/omim:210600> ;
        rdfs:comment    """This association has p-value of 0.000656, has attribute gene CENPJ (Entrenz
gene id 55835) and attribute disease Seckel Syndrome (OMIM 210600)."""@en .

    :Association_1_p_value a sio:probability-value ;
        sio:has-value   "0.000656"^^xsd:float .
}
```
**Figure 11 Nanopublication assertion, adapted from [40].** Statistical association between gene and disease expressed using Bio2RDF and SIO ontology. [41,42].

The W3C Provenance working group later released the PROV-O Ontology [3] as a Recommendation. PROV-O provides a general way to describe provenance relations using OWL. It describes provenance as a set of interactions between *Entities*, *Agents and Activities*. Interactions can be described using direct relations like *prov:wasGeneratedBy*; or they may be described with qualified indirect relationships, using classes like *prov:Generation*. The latter allows assignment of roles to agent participations and other details such as timestamp and location.

PROV-O is a generic framework for describing provenance in a whole range of applications, but used alone it lacks necessary detail for the more specific provenance of authoring and versioning that arise from our use cases. PAV can be a useful specialization of PROV-O by providing simple relationships for expressing common provenance for digital artifacts. Therefore PAV, starting from version 2.1, introduced a mapping from PAV to PROV-O using subproperties, as detailed in Table 4:

In PROV-O, entities are considered immutable, and different states for the purposes of provenance are represented as different entities, each with their own provenance. The W3C Provenance Working Group has published a note "Dublin Core to PROV Mapping" [47], proposing a subproperty mapping from Dublin Core Terms to PROV-O, in addition to a "complex mapping" by using SPARQL CONSTRUCT to create detailed PROV-O traces. This note highlights the difference between Dublin Core and PROV-O resources: while the former conflates more than one version or "state" of the resource in a single entity, the latter proposes to separate all of them.

We have provided a subproperty mapping from PAV to PROV-O, which implies a similar entity conflation, by attaching all properties to the same resource rather than introducing intermediate entities, which would be required to give a detailed PROV-O trace of the activities that lead to the generation of the final resource state. The combination of OWL/RDFS reasoning and the PROV inference rules [48] means we can infer further PROV statements such as entity generation and activity association from a single PAV statement, shown in Figure 13.

Note that some information is not preserved in these inferred statements, e.g. the distinction between authors and curators; and the PAV statements of authorship and authorship time are detangled into separate existential variables and PROV-O statements.

A more integrated mapping from PAV properties to such chains of PROV-O activities and entities could be formulated in a similar fashion to that shown in [47],

```
:PublicationInfo {
    <#> dcterms:created "2012-10-26T11:32:30.758274Z"^^xsd:dateTime ;
    pav:authoredBy <http://www.researcherid.com/rid/B-6035-2012> ,
                   <http://www.researcherid.com/rid/B-5927-2012> ;
    pav:createdBy  <http://www.researcherid.com/rid/B-5852-2012> ;
    pav:version "1.4" ;
    dcterms:hasVersion <http://rdf.biosemantics.org/vocabularies/gene_disease_nanopub_example_version4#>;
    dcterms:rights <http://creativecommons.org/licenses/by/3.0/> ;
    dcterms:rightsHolder <http://biosemantics.org> .
}
```
**Figure 12 Nanopublication attribution specifying attribution information of the nanopublication in Figures 10 and 11.** PAV is used to distinguish between the author of the nanopublication (the scientists who made the assertion expressed in Figure 10), and the creator of its digital representation, who in this case expressed the assertion as an RDF graph. The nanopublication is given a *pav:version*, also identified using *dcterms:hasVersion*. (The original RDF uses the PAV 1.2 term "*versionNumber*" which was renamed to "*version*" in PAV 2.0.).

Table 4 Mapping from PAV to PROV-O

| PROV-O superproperty | PAV property | Rationale |
| --- | --- | --- |
| prov:wasAttributedTo | pav:createdBy | The creator agent participated in some activity that generated the entity. |
| | pav:createdWith | The software agent participated in some activity that generated the entity. |
| | pav:contributedBy | The contributor participated in some activity that generated the entity. |
| | pav:authoredBy | The author participated in some activity that generated the entity. |
| | pav:curatedBy | The curator participated in some activity that generated the entity. |
| | pav:importedBy | The agent (usually software in this case) participated in some import activity, which generated the entity. |
| | pav:retrievedBy | The agent (usually software in this case) participated in some retrieval activity, which generated the entity. |
| prov:wasDerivedFrom & prov:alternateOf | pav:importedFrom | Import is a transformation of an entity into another. As the resulting entity is presenting aspects of the same thing, it is also an *prov:alternateOf* the original. |
| | pav:retrievedFrom | Retrieval is construction of an entity into another. As the resulting entity is essentially (bytewise) the same, i.e. *presenting aspects of the same thing*, it is also an *prov:alternateOf* the original. Some aspects of the original entity (like its *dct:publisher*) might not be shared, and therefore *prov:specializationOf* is not an appropriate superproperty. |
| prov:wasDerivedFrom | pav:derivedFrom | Derivation is *an update of an entity, resulting of a new one*. Note that *pav:derivedFrom* is more specific than *prov:wasDerivedFrom*, and does not cover "minor" derivations as with *pav:importedFrom* and *pav:retrievedFrom*. PAV derivation implies that additional knowledge has been contributed, curated or authored. |
| prov:wasRevisionOf | pav:previousVersion | The new version is a *revised version of the original*. *pav:previousVersion* is more specific than *prov:wasRevisionOf* because it is intended for minor updates and corrections, and typically would be used with linearly incremental *pav:version* numbers. Significant changes (contributing new knowledge) should be indicated with *pav:derivedFrom*. |
| prov:wasInfluencedBy | pav:sourceAccessedAt | The source Entity has an *effect on the character, or development of the entity*. The PAV term is a weak indication that another resource was consulted (for instance as part of curation), but without the new entity being directly derived from the source. The *prov:hadPrimarySource* is not an appropriate superproperty, as it implies a stronger statement, giving the source the status of a *primary source* and derivation. As the resource is not necessarily based on or transformed from the consulted source, we can't assume *prov:wasDerivedFrom* as a superproperty. |

which explores complex mappings of Dublin Core Terms to detailed PROV-O patterns. For instance, unrolling PAV import statements to PROV-O activities could create triples as shown in Figure 14.

Detailing such a mapping is currently work in progress, and would have to balance logical correctness vs. usefulness, for instance the above assumes that all PAV import statements describe the same activity, but if there are multiple *pav:importedFrom* statements and multiple *pav:importedBy* statements we cannot be certain about the extent of that import activity.

The authors believe that the current PROV-O sub-property mapping is liberal enough to allow PAV to complement more detailed provenance traces using PROV-O, while enabling inferences to compatible PROV-O statements.

In order to demonstrate PAV interoperability with PROV-O, we wanted to make use of the PROV-O mapping, so that PROV-O statements could be inferred from PAV statements using a standard OWL reasoner. We then wanted to test if a PROV-O consuming tool was able to understand the statements.

```
                          <http://example.com/resource.html> pav:authoredBy :paolo .

       (subproperty) →    <http://example.com/resource.html> prov:wasAttributedTo :paolo .

(attribution-inference) → <http://example.com/resource.html> prov:wasGeneratedBy _:activity .
                          :paolo prov:wasAssociatedWith _:activity .
```

**Figure 13 Inferences from PAV authorship to existential PROV-O activities.** *pav:authoredBy* is subproperty of *prov:wasAttributedBy*, which, according to PROV constraint attribution-inference imply that there existed some _:activity that generated the resource and which :paolo was associated with.

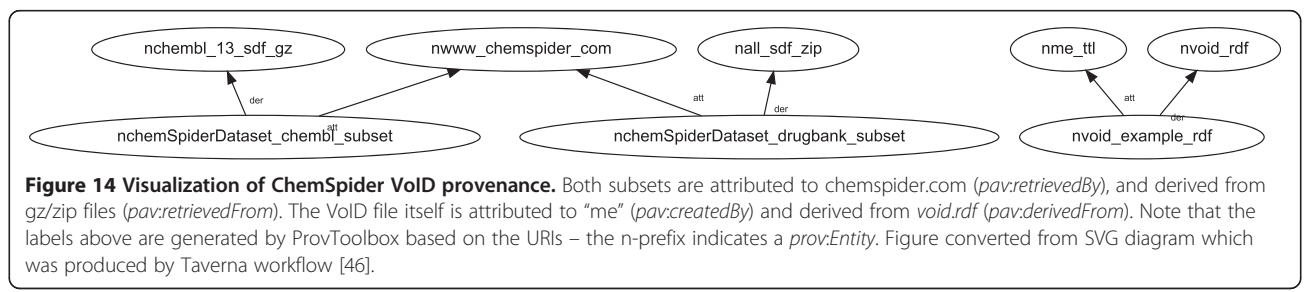

**Figure 14 Visualization of ChemSpider VoID provenance.** Both subsets are attributed to chemspider.com (*pav:retrievedBy*), and derived from gz/zip files (*pav:retrievedFrom*). The VoID file itself is attributed to "me" (*pav:createdBy*) and derived from *void.rdf* (*pav:derivedFrom*). Note that the labels above are generated by ProvToolbox based on the URIs – the n-prefix indicates a *prov:Entity*. Figure converted from SVG diagram which was produced by Taverna workflow [46].

To perform this experiment, we used Taverna Workbench v2.4 [49] to build a workflow [50] using the OWL reasoner Pellet v2.3.0 [51], and then visualized the inferred statements as a diagram (SVG) by using ProvToolbox v0.1.2 [52], which understands PROV-O. We executed this workflow against the VoID example [53] in the Open PHACTS dataset specification [28] to visualize its PAV statements. The generated provenance diagram is shown in Figure 14.

## Method

The PAV ontology was developed with the aim of enabling traceability of scientific results and their representations. The design was driven by real requirements, which initially stemmed from the AlzSWAN project, and later were expanded to requirements from other projects. DC terms and Open Provenance Model [4] were available at that time, and could have been used. However, they were found not to be suitable. Specifically, DC Terms conflates different agent roles that the participants in the SWAN project want to distinguish, in particular, authorship, creation and contribution. OPM adopts a process-oriented view of lineage by detailing the processes whereby artifacts are used and generated, and the agents that controlled those processes. Instead, the requirements elicited in the context of the SWAN project, and other projects later on, targeted mainly the expression of lineage in terms of the relationships between the artifacts, and the relationships between the artifacts and the agents involved in their creation, authorship or curation.

For these reasons, we decided to develop a new vocabulary that specifically addressed the SWAN project requirements. In doing so, the following principles were followed:

*Keep the ontology lightweight*: Experience suggests that a complex and large vocabulary (albeit well crafted) is likely not to be adopted by users. Therefore, the authors of PAV were keen to have a minimal set of terms (properties) that cater for the needs identified in the context of the AlzSWAN project. In particular, PAV does not attempt to model the complete chain of process-oriented provenance.

*Favoring Incremental (and organic) development*: Rather than trying to design an ontology and convince the users to utilize it, the development of the PAV ontology went through cycles in which the ontology designers communicated with end users and examined how the ontology is used in practice. Modifications and additions were then made based on the observations made in each cycle.

*Reuse and recommend existing vocabularies when they cater for given requirements*: For those requirements supported by existing vocabularies, the authors of the PAV ontology strived to either recommend their reuse or (when necessary) extend them. In this respect, we have shown in the previous section, how PAV extends terms from the PROV-O ontology.

The PAV ontology is currently in its second version. Since its inception, PAV has gained momentum, and it is now adopted by several vocabularies and projects. It is increasingly viewed as one of the main vocabularies for specifying provenance information in the biomedical semantics field. Some of the applications and projects that have adopted PAV are indicated in the Results section. In 2009, PAV was used as one of its starting vocabularies by the W3C Provenance Incubator group. The Incubator group preceded the W3C Provenance Working Group, which made the current PROV specifications.

## Discussion

In this section, we analyze and compare three proposals we have found to be relevant to PAV: Dublin Core Terms, BIBFRAME and Provenance Vocabulary (PRV). We then close the article by presenting some concluding remarks.

### Dublin core terms

The Dublin Core Terms vocabulary provide terms such as *dct:contributor* and its subproperty *dct:creator*, and we have argued that they conflate the roles of content authoring, knowledge curation and representation creation. Although our presented use-cases highlights the importance of distinguishing these in the setting of formal knowledge representation, the ambiguous definition of *dct:creator* also means that its value for stating consistent provenance is significantly reduced on the web in general. In cases where the content author and representation creator are different, common use of DC Terms [2] is often to attribute only one of these, but which one depends on the application. A use case could be a corporate blog where a webmaster (Bob) types in an announcement, which the CEO (Alice) sent in an email. Some blog platforms would automatically represent the currently logged in user (Bob) as the *dct:creator*, other platforms might allow the webmaster to select an author (Alice) from the corporate directory and would instead represent her as the *dct:creator*.

Using PAV, the blog platform can be more precise about the provenance of the post. When the platform has no user interface for describing the author, the safest would be to present *pav:createdBy* for the current user. If the user interface allows selecting a different author, then both *pav:authoredBy* and *pav:createdBy* can be supplied. Enterprise publishing platforms could also indicate curation (e.g. hyperlinks and textual formatting) with *pav:curatedBy* and additional contributions (such as adding an illustration) with *pav:contributedBy*.

Dublin Core Terms defines terms that may cover some provenance aspects (*dct:isFormatOf*, *dct:source*, *dct:isVersionOf*, *dct:replaces*), however DC Terms concerns itself primarily with catalogue metadata for a resource, while PAV has a bigger focus on entity-agent-driven provenance.

For instance, *dct:isFormatOf* is an existential statement that there is a different representation of the same content, while *pav:importedFrom* also implies directionality and a transformation step which was performed by an agent, indicated with *pav:importedBy*. The former term is useful for finding alternate representations, while the latter PAV relation gives lineage to the resource, which can be beneficial for instance to track down the source of an inconsistency or to verify that data is current.

As *dct:creator* can be seen to cover both content authoring and creating its representation, we have defined both *pav:authoredBy* and *pav:createdBy* to be sub-properties of *dct:creator*, while its superproperty *dct:contributor* is a superproperty of *pav:contributedBy*; here the PAV term only covers contributions to the work or content, while *dct:contributor* may also cover representational contributions such as scaling an image or converting HTML to PDF.

Other PAV properties have not been mapped to DC Terms in the OWL ontology. Part of the reason for this is that the DC Terms vocabulary is not fully OWL compatible (e.g. properties are not declared as either annotation or object properties), another is that we found the more bibliographic DC Terms to be hard to align with the provenance oriented aspect of PAV using strict OWL property hierarchies.

In order to clarify the differences between the remaining properties that might seem similar between PAV and Dublin Core Terms and to relate the two vocabularies in detail, we defined a SKOS mapping [54]. As the differences are often conceptual we found the use of SKOS [43] more beneficial than a formal OWL mapping. The most significant mappings are shown in Table 5 with their rationale:

### BIBFRAME

The Library of Congress officially launched its Bibliographic Framework Transition Initiative (BIBFRAME) [55,56] initiative in May 2011. The initiative aims to re-envision the current standard for bibliographic exchange (MARC 21) [57] and implement a new bibliographic environment for libraries that makes "the network" central and interconnectedness commonplace, using semantic web technologies. As PAV can be used to express attribution of both digital resources and traditional publications, and BIBFRAME is an emerging standard within the library community, we here explore BIBFRAME and compare it with PAV.

BIBFRAME revolves around two main concepts: the *Creative Work*; and the *Instance*, reflecting an individual, material embodiment of the Work. These are similar distinctions to the *Work* and *Manifestation* dichotomy in the original Functional Requirements for Bibliographic Records (FRBR) [58] model, or in other related models such as the FRBR-aligned Bibliographic Ontology (FaBiO) [59]. An example of the BIBFRAME two-level model is depicted in Figure 15.

Although PAV itself does not distinguish between work and instances, the distinction between content and its representation is at the core of PAV; exemplified by *pav:authoredBy* vs. *pav:createdBy*. There is however nothing inherent with PAV itself that prevents its usage with separate Work and Instance resources. In fact, the PROV-O property *prov:specializationOf* is intended for modeling abstraction levels, so if a *bf:Work* is *pav:authoredBy* Alice, and a *bf:Instance* is a *prov:specializationOf* the work, then the instance can be implied to also be *pav:authoredBy* Alice.

For bibliographic data, multiple abstractions levels such as in BIBFRAME and FRBR are elegant and useful, but for many other use cases, such as for provenance of a blog post or nanopublication, the separation of *instance* and *work* can be inconvenient, hard or even impossible to achieve. PAV, as a general vocabulary for provenance and authoring of resources, is applicable in both approaches.

### Provenance vocabulary

While PAV allows expression of data sources (*pav:sourceAccessedAt*) and derivations (*pav:derivedFrom*, *pav:importedFrom*), the Provenance Vocabulary (PRV) [5,61] is an extension of PROV-O to express more detailed provenance of data items on the web, by forming chains of *prov:Activity* to detail data creation, retrieval, access and publication. Each activity can be associated with *prov:Agents* which directly or indirectly perform the work. Additional PRV modules allow expression of database queries, HTTP retrieval and file operations, and therefore PRV might at first glance seem like an alternative to PAV. We here consider a use case where Steiner et al. adopted PRV. Below, we explore the complexity of querying the process-oriented PRV approach and we demonstrate how PAV can complement such detailed provenance and simplify queries.

**Table 5 SKOS mappings of applicable PAV terms to Dublin core terms**

| SKOS mapping | Rationale |
|---|---|
| **pav:authoredBy skos:broadMatch dct:creator** | Broad match due to the common usage of *dct:creator* to mean the creator of the Work rather than just the creator of the particular resource. Solely creating the representation of a resource is in PAV covered instead by *pav:createdBy*, but would often also be covered by *dct:creator*, therefore this is not a *skos:closeMatch*. |
| **pav:contributedBy skos:closeMatch dct:contributor** | Close match due its the common usage to mean someone who added to the Work of the resource (usually not just the digital representation), but not *skos:exactMatch* as purely representational contributions represented with *dct:contributor* should be mapped to *pav:createdBy*. |
| **pav:createdBy skos:broadMatch dct:creator** | A PAV creator is a particular kind of DC Terms creator, which made the digital representation of the resource. |
| **pav:importedFrom skos:broadMatch dct:source** | Imported is a specialization of being derived from the related resource in whole. |
| **pav:importedFrom skos:broadMatch dct:isFormatOf** | The resulting resource is *substantially the same as the source, but in another format*. However imported also implies provenance of a directed transformation from the original, at a given time and performed by an agent, and hence this is a broad match. |
| **pav:importedBy skos:broadMatch dct:creator** | The agent importing is essentially a specialized creator of the new resource, hence has close match *dct:creator*. In common use of DC Terms the extent of the transformation work might however affect whether a *dct:creator* corresponds to an importer or author. |
| **pav:derivedFrom skos:broadMatch dct:source** | A related resource from which the described resource is derived, but *pav:derivedFrom* is more specific (*skos:broadMatch*) than *dct:source*, as it requires further contributions to the content, and does not cover say *pav:importedFrom* or *pav:retrievedFrom*. |
| **pav:derivedFrom skos:narrowMatch dct:isVersionOf** | *pav:derivedFrom* do point to a resource of which the '*described resource is a version, edition, or adaptation*', but also allow further derivations, and so has a narrow match *dct:isVersionOf*. The *pav:derivedFrom* does require such contributions to be in the form of content and not just representation, which corresponds closely to *dct:isFormatOf* 's description' Changes in version imply substantive changes in content rather than differences in format'. |
| **pav:previousVersion skos:narrowMatch dct:replaces** | *dct:replaces* is a stronger statement (*skos:narrowMatch*) than *pav:previousVersion*, as the PAV statement does not necessarily imply the original was superseded. For instance, a draft specification may be *pav:previousVersion* a previously published specification, but it is not *dct:replaces* the previous version as the draft is not official yet. |
| **pav:previousVersion skos:relatedMatch dct:isVersionOf** | *pav:previousVersion* is only considered to have a related match *dct:isVersionOf*, as *pav:previousVersion* does not generally cover 'substantive changes in content'. |

PRV was conceived in 2009, and has been adapted to describe provenance of a range of internet resources, from OpenStreetMap [62] and readings in sensor networks [63] to reified RDF statements [64] and Facebook posts [65]. Here we explore the last case, which presents a browser extension and a REST service for annotating Facebook microposts by combining several natural language processing (NLP) APIs to tag posts with semantic terms from vocabularies like dbpedia.org [66]. The service uses the Provenance Vocabulary (PRV) to indicate how the underlying text mining APIs have contributed to its tagging. This provenance is expressed in rich details of the processes of data creation and multiple data retrievals, including individual API calls, embedding details of their HTTP transactions using the HTTP Vocabulary [67]. An abbreviated example of the resulting tag is included in Figure 16.

The authors of [65] are conscious of the need to reduce the verbosity of their provenance trace, and list this as a consequence of using the PRV vocabulary. We have explored their use of the PRV model and based on their example listing, formulated how one could find out: (i) the APIs called to create the tagging *<tag1>*, (ii) when the tag was made, and (iii) which agent created the tag. In order to answer this, we have to query through the individual processes of data creation, retrieval and access, as shown in Figure 17.

Using PRV and process-oriented modeling allows the service to express such provenance in detail, but forming this query requires in-depth knowledge about the particular graph structure, which mirrors how the service creates, retrieve and access data. As such, the underlying structure might change significantly if the mechanisms of the service are modified, requiring query rewrites.

The equivalent PAV statements can be queried in a simpler way, as shown in Figure 18.

```
                @prefix bf: <http://bibframe.org/vocab/> .
                @prefix madsrdf: <http://www.loc.gov/mads/rdf/v1#> .
                @prefix rdf: <http://www.w3.org/1999/02/22-rdf-syntax-ns#> .
                @prefix sample: <http://bibframe.org/resources/sample-lc-1/> .

            sample:16300892      a bf:LanguageMaterial, bf:Work;
                bf:title "Halo evolutions : essential tales of the Halo universe";
                bf:contains sample:work36, sample:work37, sample:work38, sample:work39, sample:work40;
                bf:derivedFrom <http://id.loc.gov/resources/bibs/16300892>;
                bf:instance sample:instance41, sample:instance42;
                bf:language <http://id.loc.gov/vocabulary/languages/eng>;
                bf:subject sample:genre220, sample:topic217, sample:topic218,
                        sample:topic219, sample:work35, sample:work51 .

            sample:work36       a bf:Part, bf:Work;
                bf:creator sample:agent223, sample:agent224;
                bf:title "Beyond".

            sample:instance42 a bf:Instance;
                bf:title "Halo evolutions : essential tales of the Halo universe";
                bf:carrierType "paperback";
                bf:derivedFrom <http://id.loc.gov/resources/bibs/16300892>;
                bf:dimensions "21 cm.";
                bf:edition "1st ed.";
                bf:extent "526 p. :";
                bf:instanceOf <http://bibframe.org/resources/sample-lc-1/16300892>;
                bf:isbn10 <http://www.lookupbyisbn.com/Search/Book/0765315734/1>;
                bf:placePub <http://bibframe.org/resources/sample-lc-1/place28>;
                bf:provider <http://bibframe.org/resources/sample-lc-1/organization236> .
```

**Figure 15 Example of BIBFRAME representation of a book as a creative work (*sample:16300892*) and its paperback instance (*sample: instance42*) which have features such as dimensions and pages.** Note how this work contains parts (tales) that themselves are works, each having individual *bc:creators*, and how *sample:16300892* (the bibliographical record, not the work) is *bf:derivedFrom* another bibliographical record. Adapted from RDF/XML example at [60].

We argue that PAV can provide a simpler way to describe provenance from the perspective of the interesting resource. This allows writing general provenance queries without a pre-existing understanding of the specific mechanisms that made the resource.

This simplification does come at a small cost: If multiple *?api* resources have been imported for the same tag, it is not possible with PAV alone to express when each individual *?api* was accessed, as import details such as *?when* and *?agent* are expressed directly on the resulting resource. We believe this trade-off is reasonable as for common cases there will be a single resource for each of *pav:importedFrom*, *pav:importedOn* and *pav: importedBy*.

Our design decision to not express the implied activities is reflected in all PAV properties such as *pav: authoredOn, pav:authoredBy*; or *pav:sourceAccessedAt, pav:sourceAccessedBy, pav:sourceAccessedOn*; and this reflects the simplicity approach of PAV.

This simplicity of PAV's approach does not preclude the concurrent expression of more detailed provenance using other vocabularies such as PRV; as we showed in Figures 18 and 19, more specific details can be expressed by unrolling a PAV statement into a chain of corresponding PROV-O activities and entities. We believe PRV can be used such to compliment PAV for details (and vice versa), and as both ontologies specialize PROV-O without enforcing significant constraints, a PROV-O aware client can follow the traces across both vocabularies; although without gaining the specialized understanding expressed using PAV or PRV.

## Conclusions
In this article we have presented the PAV ontology, a lightweight vocabulary for capturing provenance, authorship and versioning of resources on the Web. PAV distinguishes between the roles of content contributors (including authors and curators) and creators of

```
                    :G = {
                    <https://www.facebook.com/Tomayac/posts/10150175940867286> ctag:tagged [
                        a ctag:Tag ;
                        ctag:label "BibTeX" ;
                        ctag:means <http://dbpedia.org/resource/BibTeX>
                    ] .
                    } .
                    :G a prv:DataItem, rdfg:Graph ;
                    prv:createdBy [ a prv:DataCreation;
                        prv:usedData [
                            prv:retrievedBy [ a prv:DataAccess ;
                                prv:accessedResource <http://spotlight.dbpedia.org/rest/annotate> ;
                                prv:exchangedHTTPMessage [ a http:Request;
                                    http:methodNAme "GET" . # ...
                                ]
                            ]
                        ] # ...
                    ] .
```

**Figure 16 Example of tagging a Facebook post with DBPedia terms using the Common Tag vocabulary [41].** The provenance of the graph that contains the ctag statement expresses a chain of PRV data creation and access activities [61]. In TriG format, abbreviated from figure in [65].

representations. Additionally PAV can describe which sources have been accessed, transformed or consumed in order to create the resource.

As well as the ontology, we have listed examples of projects that have adopted PAV, illustrating their usage through concrete examples. Furthermore, we have presented how PAV extends the W3C recommendation PROV-O, and how this enables detailed provenance traces in PROV-O to be combined with PAV's direct relationships to the origins of a resource.

Originally created in 2006 with curated knowledge bases (such as AlzSWAN) in mind, PAV has evolved and is now used to document a wide variety of digital resources. PAV introduces terms for clearly attributing the intellectual property of the content, and also deals with other aspects crucial for representing scientific content

```
CONSTRUCT {
  ?entity   prov:wasGeneratedBy    _:import ;
            prov:wasDerivedFrom    ?source .
  _:import  a                      :ImportActivity ;
            prov:used              ?source ;
            prov:wasAssociatedWith ?agent ;
            prov:endedAtTime       ?when .
} WHERE {
  ?entity   pav:importedFrom       ?source ;
            pav:importedAt         ?when ;
            pav:importedBy         ?agent .
}
```

**Figure 17 SPARQL query over PRV provenance to find data creation, retrieval and access of a Facebook tag.** The query finds the activity the tag was prv:createdBy, which was prv:performedBy the agent and prv:usedData that were prv:retrievedBy another activity, which prv:accessedResource the given REST API, prv:performedAt the given time.

```
SELECT ?api, ?when, ?agent
WHERE {
  <tag1>        prv:createdBy       ?dataCreation .

  ?dataCreation prv:usedData        ?data ;
                prv:performedBy     ?agent .

  ?data         prv:retrievedBy     ?dataAccess .

  ?dataAccess   prv:accessedResource ?api ;
                prv:performedAt     ?when .
}
```

**Figure 18 SPARQL query over PAV provenance to find data creation, retrieval and access of a Facebook tag (equivalent to Figure 16).** The tag was pav:importedFrom the given REST API, by the importing agent at the given import time.

```
SELECT ?api, ?when, ?agent
WHERE {
   <tag1> pav:importedFrom ?api ;
          pav:importedAt ?when ;
          pav:importedBy ?agent .
}
```

**Figure 19 Example of SPARQL CONSTRUCT generating PROV-O activities from PAV imports.**

in a federated environment, such as versions and resource retrieval. PAV does not specify detail about the chain of processes that produced the current state of the resource, but gives a view of attribution metadata that is uniform across a multitude of implementations.

At the core of PAV is the distinction between authoring knowledge (content) and creating representations. This is highlighted by the mapping to DC Terms, which shows how PAV properties can provide more precise attributions. Equally important, PAV derivation properties distinguish between plain retrieval, versioned updates, transformational imports, and more structural derivation. These distinctions are essential for attributing resources in the complex real world of curated knowledge bases and datasets, and PAV has been adapted for these purposes by representation models like Open PHACTS dataset descriptions, Nanopublications, Wf4Ever Research Objects and Elsevier's Satellite annotations.

We believe PAV is a good complement to existing provenance vocabularies, such as OPM and PROV-O. Indeed, use of such vocabularies often focus on describing the chain of activities that were performed to transform given resources into other resources, and the role of agents associated with those activities. In PAV, the emphasis is put on the provenance of the resources: PAV describes the lineage from other resources, and just as important, the role of agents involved with creating and maintaining the resource.

**Competing interests**
The authors declare that they have no competing interests.

**Authors' contributions**
PC conceived and authored PAV 1.0 as a spin-off of the SWAN Ontology. PC co-developed the SWAN platform and is also the main author of Annotation Ontology and the architect of the Domeo Annotation Tool. PC is one of the main authors of this article and co-developed PAV 2.x with SSR. SSR is one of the main authors of this article, and co-developed PAV 2.x with PC. SSR is one of the authors of the Research Object model together with KB, and one of the contributors of PROV-O as a member of the W3C Provenance Working Group. In this article, SSR is the author of the sections on Open PHACTS datasets, Nanopublications, PROV mapping, DC Terms and Provenance Vocabulary. He also maintains the OWL representation of PAV and the accompanying descriptions shown in the Ontology section. KB is, together with SSR, co-author of the Research Object model and the W3C PROV-O ontology. KB drafted the background and method sections, and contributed to refining the results section. He also participated in improving the descriptions of ontology terms, and by restructuring and improving the organization of the article as a whole. AG is the author of the Open PHACTS dataset descriptions, and helped validate PAV's documentation and OWL implementation. AG co-wrote the sections on Open PHACTS datasets and Nanopublications, and has helped revise the final article. CG is the Principal Investigator of the Wf4Ever project and co-investigator of the Open PHACTS project. She supervised the research and edited the final article. TC is the Principal Investigator of the SWAN project and of the Domeo Annotation Tool. He supervised the research and edited the final article. All authors read and approved the final manuscript.

**Authors' information**
Paolo Ciccarese
URL: http://orcid.org/0000-0002-5156-2703
Stian Soiland-Reyes
URL: http://orcid.org/0000-0001-9842-9718
Khalid Belhajjame
URL: http://orcid.org/0000-0001-6938-0820
Alasdair JG Gray
URL: http://orcid.org/0000-0002-5711-4872
Carole Goble
URL: http://orcid.org/0000-0003-1219-2137
Tim Clark
URL: http://orcid.org/0000-0003-4060-7360

**Acknowledgements**
We would like to thank Marco Ocana, Gwen Wong, Elizabeth Wu, and June Kinoshita for their input during the development of the SWAN ontology. We also would like to thank Paul Groth for helping us validate the mappings between PAV and the PROV model. We thank the Journal of Biomedical Semantics reviewers, who provided valuable feedback which led to further improvements of the ontology and this paper.
Stian Soiland-Reyes and Khalid Belhajjame are funded for the Wf4Ever project by the European Commission's 7th FWP FP7-ICT-2007-6 270192.
Alasdair Gray received support from the Innovative Medicines Initiative Joint Undertaking under grant agreement number 115191, resources of which are composed of financial contribution from the European Union's Seventh Framework Programme (FP7/2007- 2013) and EFPIA companies' in kind contribution.

**Author details**
[1]Department of Neurology, Massachusetts General Hospital, 55 Fruit Street, Boston, MA 02114, USA. [2]Harvard Medical School, 25 Shattuck Street, Boston, MA 02115, USA. [3]School of Computer Science, University of Manchester, Oxford Road, Manchester M13 9PL, UK.